\documentclass[aps,prl,twocolumn,preprintnumbers,superscriptaddress]{revtex4-2}
\usepackage[usenames,dvipsnames,table]{xcolor}
\usepackage{graphicx,amsmath,amssymb,amsthm,multirow,array,bm,soul}
\usepackage[mathscr]{eucal}
\usepackage[bbgreekl]{mathbbol}
\usepackage{amsfonts}
\usepackage[colorlinks,
linkcolor=blue,
anchorcolor=blue,
citecolor=red,
]{hyperref}
\usepackage{float}
\usepackage{url}
\usepackage{physics}
\usepackage{indentfirst}
\usepackage{lineno}
\usepackage{xcolor}
\usepackage{extarrows}
\usepackage{braket}
\usepackage{enumerate}

\newcommand{\llangle}{{\langle\!\langle}}
\newcommand{\rrangle}{{\rangle\!\rangle}}

\begin{document}

\title{From Weyl Anomaly to Defect Supersymmetric R\'enyi Entropy and Casimir Energy}

\author{Zi-Xiao Huang}
\email{huangzx22@m.fudan.edu.cn}
\affiliation{Department of Physics and Center for Field Theory and Particle Physics, Fudan University, Shanghai 200433, China}
\author{Ma-Ke Yuan}
\email{mkyuan19@fudan.edu.cn}
\affiliation{Department of Physics and Center for Field Theory and Particle Physics, Fudan University, Shanghai 200433, China}
\author{Yang Zhou}
\email{yang\_zhou@fudan.edu.cn}
\affiliation{Department of Physics and Center for Field Theory and Particle Physics, Fudan University, Shanghai 200433, China}

\begin{abstract}
We present a closed-form expression for the contribution of surface defects to the supersymmetric Rényi entropy in six-dimensional $(2,0)$ theories. Our results show that this defect contribution is a linear function of $1/n$ and is directly proportional to $2b-d_2$, where $b$ and $d_2$ are the surface defect Weyl anomaly coefficients. We also derive a closed-form expression for the defect contribution to the supersymmetric Casimir energy, which simplifies to $-d_2$ (up to a proportionality constant) in the chiral algebra limit.
\end{abstract}

\maketitle
\textit{Introduction.}
Defects play a pivotal role in quantum field theory, with familiar examples including Wilson lines and Wilson surfaces. These defects often represent external probes coupled to the theory, such as a charged heavy probe particle in gauge theory, which is described by a Wilson line and serves as a key tool for understanding the non-perturbative properties of the theory. Meanwhile, entanglement entropy (EE) and R\'enyi entropy have garnered significant attention in recent years for their role in linking information theory, field theory, and gravity. These quantities are closely related to conformal anomalies in conformal field theories and can be computed holographically. As a one-parameter generalization of EE, R\'enyi entropy provides deeper insights into the structure of quantum field theories, making it essential to account for defect contributions to R\'enyi entropy.

While the defect contribution to EE has been studied in previous works~\cite{Jensen:2013lxa,Lewkowycz:2013laa,Kobayashi:2018lil,Estes:2018tnu,Gentle:2015jma,Gentle:2015ruo}, a method for computing the corresponding contributions to R\'enyi entropy has remained elusive. In this letter, we take a significant step forward by calculating the contribution of surface defects to the supersymmetric R\'enyi entropy. A key motivation for studying surface defects in six dimensions is to gain insights into $(2,0)$ theories, whose proper formulation remains a challenging open problem. By combining supersymmetry with R\'enyi entropy, one can define a supersymmetric refinement of the ordinary R\'enyi entropy for these theories, as proposed in prior works~\cite{Nian:2015xky,Zhou:2015kaj}. This approach involves placing $(2,0)$ theories on $\mathbb{S}^1_\beta \times \mathbb{H}^5$ with an additional R-symmetry background field. Notably, the supersymmetric R\'enyi entropy exhibits universal relations with the conformal and 't Hooft anomalies and can be computed holographically via two-charge hyperbolic black holes in the large $N$ limit.

Building on these developments, we analyze the contribution of the most natural operator, the surface operator, in $(2,0)$ theories on $\mathbb{S}^1_\beta \times \mathbb{H}^5$. Our findings reveal that the surface operator contribution to the supersymmetric R\'enyi entropy is a linear function of $1/n$ and directly proportional to $2b - d_2$, where $b$ and $d_2$ are the Weyl anomaly coefficients associated with the surface defect. Additionally, we derive a closed-form expression for the defect contribution to the supersymmetric Casimir energy, which simplifies to $-d_2$ (up to a proportionality constant) in the chiral algebra limit.

\textit{Defects in CFTs.} Starting with a $d$-dimensional Euclidean CFT in flat space $\mathbb{R}^d$, introducing a $p$-dimensional planar conformal defect will break its conformal symmetry $SO(1, d + 1)$ to $SO(1, p + 1) \times SO(d - p)$. The CFT stress-energy tensor one-point function $\langle T_{\mu\nu} \rangle$ in flat space vanishes due to the conformal symmetry. However, in the presence of defect, this one-point function does not necessarily vanish. If the planar defect is located at $x^i=0$, we can write the ambient metric as $\text{d}s^2 = \text{d} \hat x^a \text{d} \hat x^a + \text{d}x^i \text{d}x^i$, with $a=0, \dots, p-1$ the indices labeling the directions parallel to the defect and $i=p, \dots, d-1$ for the transverse directions. Varying the defect partition function gives the defect CFT stress-energy tensor, which can be split into the ambient part $T^{\mu\nu}$ and the defect localized part $\hat{T}^{ab}$~\cite{Billo:2016cpy,Kobayashi:2018lil}. The ambient stress-energy tensor is a symmetric traceless tensor of dimension $d$ and spin $2$, and from the (partial) conservation and residual conformal symmetry, its one-point function in the presence of the defect can be fixed~\footnote{The notation $\llangle O \rrangle$ refers to the correlation function measured in the presence of defect $\llangle O \rrangle \equiv \langle O D \rangle / \langle D \rangle$, where $D$ denotes the defect. }
\begin{equation}\label{Tbulk}
\begin{split}
\llangle T^{ab} \rrangle &= -\frac{d - p - 1}{d} \frac{h}{|x^i|^d} \delta^{ab}\ , \quad \llangle T^{ai} \rrangle = 0\ ,\\
\llangle T^{ij} \rrangle &= \frac{h}{|x^i|^d} \left(\frac{p + 1}{d}\delta^{ij} - \frac{x^i x^j}{|x^i|^2}\right)\ ,
\end{split}
\end{equation}
up to a factor $h$, which characterizes the property of the defect~\cite{Billo:2016cpy}.
In this letter we are particularly interested in surface defects with $p=2$. In this case, there will be defect Weyl anomalies, similar to the Weyl anomalies in even-dimensional CFTs.

\textit{Defect Weyl anomalies.} Even-dimensional defects suffer from Weyl anomalies, i.e., the trace of the defect localized stress-energy tensor $\hat{T}$ does not vanish. For surface defects~\cite{Graham:1999pm, Henningson:1999xi, Schwimmer:2008yh}, 
\begin{equation}
    \llangle \hat{T}^a_a \rrangle = - \frac{1}{24\pi} \left[ b R^\Sigma + d_1 \Tilde{\Pi}_{ab}^{\mu} \Tilde{\Pi}^{ab}_{\mu} - d_2 W_{ab}^{ab} \right]\ , 
\end{equation}
with $R^\Sigma$ the intrinsic Ricci scalar of the defect submanifold $\Sigma$, $\Tilde{\Pi}_{ab}^{\mu}$ the traceless part of the second fundamental form, and $W_{abcd}$ the pullback of the bulk Weyl tensor. The coefficients $b$, $d_1$ and $d_2$ are defect central charges, where $d_1$ is proportional to the coefficient of the displacement operator two-point correlator and $d_2$ is related to $h$ in \eqref{Tbulk} through~\cite{Lewkowycz:2014jia, Bianchi:2015liz, Jensen:2018rxu}
\begin{equation}\label{eq-hd2}
d_2 = 6\pi\Omega_{d-3}\frac{d-1}{d}\, h\ .
\end{equation}
The Weyl anomaly of surface defect is also manifested in the logarithmic divergence of its expectation value
\begin{equation}\label{eq-int-anomalyDef}
    \log \langle D[\Sigma] \rangle \supset \int_\Sigma \mathcal{A}_\Sigma \mathrm{vol}_\Sigma \log\ell/\epsilon\ ,
\end{equation}
with the anomaly density $\mathcal{A}_\Sigma$ given by
\begin{equation}\label{eq-anomalyDef}
    \mathcal{A}_\Sigma = \frac{1}{24\pi} \left[ b R^\Sigma + d_1 \Tilde{\Pi}_{ab}^{\mu} \Tilde{\Pi}^{ab}_{\mu} - d_2 W_{ab}^{ab} \right]\ . 
\end{equation}
From \eqref{eq-int-anomalyDef} and \eqref{eq-anomalyDef}, one can see that the free energy of a spherical surface defect is~\footnote{In this letter we always focus on the defect contribution to various quantities thus we omit the superscript ``defect''. }
\begin{equation}
F = -\log\langle D[\mathbb{S}^2]\rangle = -\frac{b}{3} \log \ell/\epsilon\ .
\end{equation}
Defect Weyl anomalies also determine the defect contribution to EE $\tilde{S}$. It was shown in~\cite{Jensen:2018rxu} that the surface defect contribution to EE is~\footnote{we use $\tilde{S}$ to denote non-SUSY entropies and $S$ for SUSY ones.} 
\begin{equation}\label{eq-JensenOBannon}
\begin{split}
\tilde{S} &= -F + \beta E = \log\langle D[\mathbb{S}^2]\rangle + \int \llangle T^\tau_\tau \rrangle\\
&= \frac{1}{3}\left(b - \frac{d-3}{d-1}d_2\right)\log\ell/\epsilon\ ,
\end{split}
\end{equation}
where $\tau$ is the coordinate along the replica direction.
This formula can be used to calculate the central charges for a class of surface defects.
For 1/2-BPS Wilson surfaces in 6d $A_{N - 1}$ $\mathcal{N} = (2,0)$ SCFT, both $\llangle T^{\mu\nu}\rrangle$ and $\tilde{S}$ in \eqref{eq-JensenOBannon} can be calculated holographically~\cite{Gentle:2015jma, Estes:2018tnu}. Using these results, \eqref{Tbulk} and \eqref{eq-JensenOBannon} give
\begin{equation}\label{eq-b-d2-R}
b = 24 \left(\Lambda, \rho\right) + 3 \left(\Lambda, \Lambda\right)\ , \quad d_2 = 24 \left(\Lambda, \rho\right) + 6 \left(\Lambda, \Lambda\right)\ ,
\end{equation}
with $\Lambda$ the highest weight of the defect representation $\mathcal{R}$ in $A_{N - 1}$ Lie algebra $\mathfrak{su}(N)$, $\rho$ the Weyl vector of $\mathfrak{su}(N)$, and $(\cdot,\cdot)$ the inner product in the Lie algebra. In the case of the symmetric representation $(k)$, \eqref{eq-b-d2-R} gives
\begin{equation}\label{eq-b-d2-Sk}
\begin{split}
b_{(k)} &=  12 N k + 3 k^2 - 12 k - 3 k^2/N\ ,\\
d_{2(k)} &= 12 N k + 6 k^2 - 12 k - 6 k^2/N\ .
\end{split}
\end{equation}
While for the anti-symmetric representation $[k]$, 
\begin{equation}\label{eq-b-d2-Ak}
\begin{split}
b_{[k]} &= 12 N k - 12 k^2 + 3 k - 3 k^2/N\ ,\\
d_{2[k]} &= 12 N k - 12 k^2 + 6 k - 6 k^2/N\ .
\end{split}
\end{equation}
The main purpose of this letter is to show that, for 1/2-BPS Wilson surfaces in 6d $(2,0)$ theories, the defect contribution to supersymmetric R\'enyi entropy is fixed by $b$ and $d_2$. 

\textit{Supersymmetric R\'enyi entropy.} 
R\'enyi entropy is a one parameter generalization of EE and provides information about the entanglement spectrum. It appears when we compute EE using the replica trick and returns to EE in the limit $n\to 1$,
\begin{equation}\label{eq-RE}
S_n \equiv \frac{1}{1-n} \log\mathrm{Tr}\rho^n = \frac{1}{1-n} \log\frac{Z_n}{Z^n}\ ,
\end{equation}
where $n$ is the R\'enyi index and $Z_n$ is the partition function on the $n$-replica space. 
For a $d$-dimensional CFT in flat space, the $n$-replica space with a spherical entangling surface can be mapped to $\mathbb{S}_\beta^1\times \mathbb{H}^{d - 1}$ with $\beta = 2\pi n $ using conformal transformations~\cite{Casini:2011kv, Klebanov:2011uf}. In this letter we mainly focus on 6d $(2,0)$ theories with $d=6$. The replica trick generally breaks the supersymmetry because of the conic singularity. Namely, there will be no surviving Killing spinors for $n\neq 1$~\cite{Nian:2015xky}. To preserve supersymmetry, one should turn on an extra R-symmetry background fields, leading to the observable of \textit{supersymmetric R\'{e}nyi entropy (SRE)}~\cite{,,Nian:2015xky,Zhou:2015kaj}.
The R-symmetry group of 6d $(2,0)$ theories is $SO(5)$ that has two $U(1)$ Cartans, and therefore one can turn on two independent R-symmetry background gauge fields (chemical potentials) to twist the boundary conditions along the replica circle $\mathbb{S}_\beta^1$. A general analysis of the Killing spinor equation on the conic space ($\mathbb{S}^6_n$ or $\mathbb{S}_{\beta = 2\pi
n}^1 \times \mathbb{H}^5$) leads to the solution of the R-symmetry chemical potential~\cite{Nian:2015xky}~\footnote{The Killing spinors on round sphere have been explored in~\cite{Lu:1998nu}.}
\begin{equation}\label{kschemical}
\mu(n) := q_i A^i = \frac{n - 1}{2}\ ,
\end{equation}
with $q_1=q_2=1/2$ the R-charges of the Killing spinor under the two $U(1)$ Cartans. A general background satisfying \eqref{kschemical} can be expressed as~\cite{Zhou:2015kaj}
\begin{equation}\label{chemicalback}
\quad A^1 = (n-1)\, r_1\ ,\quad A^2 = (n-1)\, r_2\ , 
\end{equation}
with $r_1+r_2=1$. Assuming SRE is a polynomial of $1/n$, which is verified by free field calculations as well as large $N$ results, it has been shown that the SRE for 6d (2,0) theories enjoys universal relations with Weyl anomalies (as well as ’t Hooft anomalies)~\cite{Zhou:2015kaj}. The goal of this letter is to show that the defect contribution to SRE also enjoys universal relations with defect Weyl anomalies.

\textit{Summary of the results.} The main result of this letter is the exact contribution of $\mathcal{N}=(4,4)$ surface defect to the supersymmetric R\'enyi entropy (defect SRE) of 6d $(2,0)$ theories. We show that for theories characterized by A-type Lie algebra $\mathfrak{g}$, it is a linear function of $\gamma := 1/n$
\begin{equation}\label{eq-main-result}
S_\gamma [\mathfrak{g}] = \frac{2b_{\mathfrak{g}} - d_{2\mathfrak{g}}}{6}\left[\frac{r_1 r_2}{2} (\gamma - 1) + 1 \right]\log \ell/\epsilon\ ,
\end{equation}
for any defect representation $\mathcal{R}$ in $\mathfrak{g}$, where $b_{\mathfrak{g}}$ and $d_{2\mathfrak{g}}$ are given by \eqref{eq-b-d2-R} and $r_{1,2}$ are background parameters denoting the weights of the two $U(1)$ R-symmetry chemical potentials, satisfying the constraint $r_1 + r_2 = 1$. The basic ingredients of our argument are the following: 
\begin{enumerate}[(A)]
\item At large $N$ limit, $S_\gamma$ can be calculated holographically. The result for surface defect in the fundamental representation takes the form \eqref{eq-holographyFinal}
\begin{equation}
S_\gamma [A_{N \to \infty}] = N \big( r_1 r_2 (\gamma - 1) + 2 \big)\log \ell/\epsilon\ .
\end{equation}
\item In the large $\gamma$ limit, the behavior of the surface defect SRE is controlled by the ``surface defect contribution to supersymmetric Casimir energy (defect SCE)'', which will be computed from localization as well as anomaly polynomial in (D). In particular, the result from anomaly polynomial is valid for all $N$ and all representations. Both results show that in the large $\gamma$ limit,
\begin{equation}
S_{\gamma\to \infty} \sim \gamma \log \ell/\epsilon\ .
\end{equation}
\item From (A) and (B), it is tempting to assume that defect SRE is a linear function of $\gamma$. 
\item The defect SCE for $\mathfrak{g} = A_{N - 1}$ is computed from localization \eqref{eq-logW-Sk} and \eqref{eq-logW-Ak}, and further justified using anomaly polynomial \eqref{eq-Casimir-Final}. These results give \eqref{eq-leading-coe}
\begin{equation}
\lim_{\gamma \to \infty} \frac{S_\gamma [\mathfrak{g}]}{\gamma} = \frac{r_1 r_2}{12} \left(2b_{\mathfrak{g}} - d_{2\mathfrak{g}}\right)\log \ell/\epsilon\ , 
\end{equation}
which is consistent with the large $N$ result in (A). 
\item The value of $S_\gamma$ at $\gamma = 1$ is the surface defect contribution to supersymmetric entanglement entropy (defect SUSY EE). Notice that to define EE we have to specify the boundary condition at the entangling surface. For the defect contribution, the supersymmetric boundary condition gives a different result from (\ref{eq-JensenOBannon}). We assume that it is still a linear combination of $b$ and $d_2$, which can be fixed by fitting to the holographic result in (A),
\begin{equation}
S = \frac{2b -d_2}{6}\log\ell/\epsilon\ .
\end{equation}
We also verify the defect SUSY EE as a linear combination of $\log\langle W\rangle$ and $h$ for Wilson loop in 4d $\mathcal{N}=4$ SYM, which is the 4d analogy of 6d surface defect.
\end{enumerate}
From (A)(B)(C)(D)(E) listed above, we can uniquely fix the general expression of $S_\gamma$ given in \eqref{eq-main-result}. We also verify the defect SRE \eqref{eq-main-result} directly from a modified defect SCE by using a conjecture in~\cite{Yankielowicz:2017xkf}.

\textit{Defects in $\mathbb{S}^1\times \mathbb{S}^5$ from Localization.} 
Due to the lack of Lagrangian description, it is unclear how to compute the partition function of 6d (2,0) theories directly. 
The circular reduction of 6d $A_{N - 1}$ (2,0) theories on $\mathbb{S}^1_\beta$ with an appropriate twist gives 5d $\mathcal{N}=2$ $SU(N)$ SYM theory with the coupling $g^2_{\text{YM}}=2\pi\beta$. It is therefore tempting to conjecture that the full 5d partition function on $\mathbb{S}^5$, including both the perturbative and non-perturbative parts, will capture the 6d partition function on $\mathbb{S}^1_\beta\times \mathbb{S}^5$~\cite{Douglas:2010iu, Lambert:2010iw, Kim:2012ava, Kim:2012qf}. 
Meanwhile, Wilson surfaces wrapping $\mathbb{S}^1_\beta\times \mathbb{S}^1\subset \mathbb{S}^1_\beta\times \mathbb{S}^5$ in 6d corresponds to Wilson loops in 5d. It was argued in~\cite{Bullimore:2014upa} that the 5d partition function with line defects by localization is sufficient to compute the 6d partition function with surface defects. We will adopt the same working assumption. In the strong coupling limit $\beta\to\infty$, the 5d perturbative partition function $\mathcal{Z}$ is computed by supersymmetric localization~\cite{Bobev:2015kza} and reduces to a matrix integral
\begin{equation}\label{eq-matrixI}
\int \prod_{i = 1}^N \mathrm{d}\nu_i \exp\bigg[\frac{2\pi}{\omega_1 \omega_2 \omega_3}\Big(-\frac{\pi}{\beta} \sum_{i = 1}^N \nu_i^2 + \frac{\sigma_1 \sigma_2}{2} \sum_{j < i}(\nu_i - \nu_j)\Big)\bigg]\ ,
\end{equation}
where $\nu_i$ are the scalar eigenvalues with $\nu_i<\nu_j$ for $i<j$, $\omega_{1,2,3}$ are the chemical potentials for angular momentum $J_{1,2,3}$ and $\sigma_{1,2}$ are the chemical potentials for R-charge $R_{1,2}$~\footnote{The convention for $\omega_{1,2,3}$ and $\sigma_{1,2}$ follows~\cite{Bobev:2015kza}. }. 
The chemical potentials satisfy the supersymmetric condition $\sigma_1 + \sigma_2 = \omega_1 + \omega_2 + \omega_3$. One can evaluate the integral (\ref{eq-matrixI}) using the saddle point approximation.
The saddle point solution of $\nu_i$ is 
\begin{equation}\label{eq-vi-noWL}
\begin{split}
- \frac{2\pi}{\beta} \nu_i + \frac{\sigma_1 \sigma_2}{2} \big((i - 1) - (N - i)\big) = 0\\
\Rightarrow \nu_i = \frac{\beta \sigma_1 \sigma_2}{4\pi} (2i - N - 1)\ ,
\end{split}
\end{equation}
which gives the 6d supersymmetric Casimir energy~\cite{Bobev:2015kza}
\begin{equation}
\beta E_c = -\log\mathcal{Z} = -\beta \frac{N(N^2 -1) \sigma_1^2 \sigma_2^2}{24 \omega_1 \omega_2 \omega_3}\ .
\end{equation}
The Wilson surface on $\mathbb{S}^1_\beta\times \mathbb{S}^1_j$ in 6d is expected to be captured by the 5d Wilson loop inserted on $\mathbb{S}^1_j$. When the Wilson loop is inserted along $\mathbb{S}^1_j$ with length $2\pi/\omega_j$, the localization saddles enforce the Wilson loop to make a classical contribution $\mathrm{Tr}_{\mathcal{R}} \exp\left[2\pi \nu/\omega_j\right]$~\cite{Bullimore:2014upa}. Without loss of generality, we choose $j=1$.
The trace in symmetric representation $(k)$ is
\begin{equation}\label{eq-tr-Sk}
\mathrm{Tr}_{(k)} e^{2\pi\nu/\omega_1} = \sum_{1 \leq i_1 \leq \cdots \leq i_k \leq N} \exp\left[ \frac{2\pi}{\omega_1} \sum_{l = 1}^k \nu_{i_l}\right]\ . 
\end{equation}
In \eqref{eq-tr-Sk}, the largest contribution in the summation comes with $\nu_{i_l} = \nu_N$ for all $l$~\cite{Mori:2014tca}. Therefore, the leading contribution to $\langle W_{(k)} \rangle$ is
\begin{widetext}
\begin{equation}\label{eq-matrix-Sk-WL}
\langle W_{(k)} \rangle = \frac{1}{\mathcal{Z}}\int \prod_{i = 1}^N \mathrm{d}\nu_i \exp\left[\frac{2\pi}{\omega_1 \omega_2 \omega_3}\left(-\frac{\pi}{\beta} \sum_{i = 1}^N \nu_i^2 + \frac{\sigma_1 \sigma_2}{2} \sum_{j < i}(\nu_i - \nu_j) + \omega_2 \omega_3 k \nu_N\right)\right]\ .
\end{equation}
\end{widetext}
With this approximation, the Wilson loop insertion only changes the saddle point solution of $\nu_N$, while all the other $\nu_{i < N}$ remain the same as \eqref{eq-vi-noWL}. The new saddle point solution of $\nu_N$ is 
\begin{equation}\label{eq-vN-Sk}
\begin{split}
- \frac{2\pi}{\beta} \nu_N + \frac{\sigma_1 \sigma_2}{2} (N - 1) + \omega_2 \omega_3 k = 0\\
\Rightarrow \nu_N = \frac{\beta}{4\pi} \big((N - 1) \sigma_1 \sigma_2 + 2 k \omega_2 \omega_3\big)\ .
\end{split}
\end{equation}
Evaluating the matrix integral  \eqref{eq-matrix-Sk-WL} at the saddles \eqref{eq-vN-Sk} of  $\nu_N$ and \eqref{eq-vi-noWL} of $\nu_{i < N}$ gives the defect SCE
\begin{equation}\label{eq-logW-Sk}
\begin{split}
\beta E_{(k)} &= -\log\langle W_{(k)} \rangle\\
&= -\frac{\beta}{2\omega_1} \big(k (N - 1) \sigma_1 \sigma_2 + k^2 \omega_2 \omega_3\big)\ .
\end{split}
\end{equation}
Using the same method one can evaluate the result for anti-symmetric representation~\footnote{see the supplementary material for details. }
\begin{equation}\label{eq-logW-Ak}
\beta E_{[k]} = -\frac{\beta}{2\omega_1} \big(k (N - k) \sigma_1 \sigma_2 + k \omega_2 \omega_3\big)\ .
\end{equation}

\textit{Connection between defect SRE and defect SCE.}
 Here we make the connection between the asymptotic defect SRE $S_{n\to 0}$ and the defect SCE \eqref{eq-logW-Sk} and \eqref{eq-logW-Ak}. The defect SCE of a surface defect wrapping $  \mathbb{S}^1_\beta\times\mathbb{S}_n^1 \subset \mathbb{S}^1_\beta\times\mathbb{S}_n^5  $ is related to the defect expectation value in the limit $\beta\to \infty$, i.e.,
\begin{equation}
E = -\lim_{\beta\to \infty} \partial_\beta \log \langle D_\mathfrak{g} \rangle_\beta\ .
\end{equation}
Since the defect is wrapped on $\mathbb{S}_n^1\subset\mathbb{S}_n^5$, the shape parameters should be identified as
\begin{equation}
\omega_1 = 1/n\ , \quad \omega_2 = \omega_3 = 1\ .
\end{equation}
In the limit $n\to 0$, to match the backgrounds \eqref{chemicalback}, the behavior of local chemical potentials $\sigma_1$ and $\sigma_2$ should be set as~\cite{Zhou:2015kaj}
\begin{equation}\label{eq-sigma12-n}
\sigma_1^2(n\to 0) = \frac{r_1^2}{n^2}\ , \quad \sigma_2^2(n\to 0) = \frac{r_2^2}{n^2}\ , 
\end{equation}
with $r_1 + r_2 = 1$. Evaluating \eqref{eq-logW-Sk} and \eqref{eq-logW-Ak} under the above parameter setting gives
\begin{equation}\label{eq-limit-Ec}
\begin{split}
E_{(k)}\big|_{n\to 0} &= -\frac{k(N - 1)}{2}\frac{r_1 r_2}{n}\ ,\\
E_{[k]}\big|_{n\to 0} &= -\frac{k(N - k)}{2}\frac{r_1 r_2}{n}\ .
\end{split}
\end{equation}
Based on \eqref{eq-limit-Ec}, \eqref{eq-b-d2-Sk} and \eqref{eq-b-d2-Ak}, we find that
\begin{equation}\label{SCEbd}
E\big|_{n\to 0} = -\frac{2b - d_{2}}{24}\frac{r_1 r_2}{n}\ ,
\end{equation}
which can be further justified using anomaly polynomial. In fact, from the derivation by anomaly polynomial in the next section, we will see that the relation (\ref{SCEbd}) holds for all $N$ and all defect representations.
It has been observed in~\cite{Zhou:2015kaj,Yankielowicz:2017xkf} that the $n\to 0$ limit of SRE is controlled by the extremally squashed SCE up to a factor. Therefore one can take use of the latter to determine the large $\gamma$ limit of SRE. Here we use the same idea to determine the large $\gamma$ behavior of defect SRE.
The defect free energy at $\beta\to\infty$ and $n\to 0$ is
\begin{equation}
F[\mathbb{S}^1_{n\to 0} \times \mathbb{S}^1_\beta] = \beta E_{\mathfrak{g}}|_{n\to 0} = -\beta\frac{2b_\mathfrak{g} - d_{2\mathfrak{g}}}{24}\frac{r_1 r_2}{n}\ .
\end{equation}
When $\beta \to \infty$, the defect free energy on $\mathbb{S}^1_{n\to 0} \times \mathbb{H}^1$ differs only by a volume
\begin{equation}
 F[\mathbb{S}^1_{n\to 0} \times \mathbb{H}^1] = F[\mathbb{S}^1_{n\to 0} \times \mathbb{S}^1_\beta]\frac{\text{Vol}[\mathbb{H}^1]}{\text{Vol}[\mathbb{S}^1_\beta]}\ ,
\end{equation}
from which we can obtain the asymptotic defect SRE 
\begin{equation}\label{eq-leading-coe}
\begin{split}
S_{n\to 0}[\mathfrak{g}] &=-F[\mathbb{S}^1_{n\to 0} \times \mathbb{H}^1] \\
&= \frac{2b_\mathfrak{g} - d_{2\mathfrak{g}}}{12}\frac{r_1 r_2}{n} \log \ell/\epsilon\ . 
\end{split}
\end{equation}

\textit{Defect SCE and Anomaly polynomial.} It is conjectured in~\cite{Bobev:2015kza} that for a $d$-dimensional superconformal field theory ($d$ is even), its
SCE on a space with topology $\mathcal{M} = \mathbb{S}^1 \times \mathbb{S}^{d - 1}$ is given by the equivariant integral of the anomaly polynomial, $I_{d + 2}$, 
\begin{equation}\label{eq-SCE-AP}
E_d = -\frac{1}{(2\pi)^{d/2}}\int I_{d + 2}(\mathcal{M})\ .
\end{equation}
A 2d defect wrapped on $\Sigma \hookrightarrow \mathcal{M}$ deforms the anomaly polynomial by a defect localized term
\begin{equation}
I_{d + 2}(\mathcal{M}) \rightarrow I_{d + 2}(\mathcal{M}) + \delta_{\Sigma}I_4(\Sigma)\ ,
\end{equation}
which indicates a defect version of the relation \eqref{eq-SCE-AP}. In this section, we use this idea to calculate the surface defect contribution to SCE from the equivariant integral of the defect anomaly polynomial. 

The anomaly polynomial of a 2d $\mathcal{N}=(4,4)$ surface defect in the 6d $\mathcal{N}=(2,0)$ SCFT labelled by ADE Lie algebra $\mathfrak{g}$ is given in \cite{Shimizu:2016lbw,Wang:2020xkc}
\begin{equation}
\label{eq-2d-anomaly-polynomial}
\begin{split}
I_4 =&\,\frac{1}{2} (\Lambda,\Lambda) \big(c_2(F_L)-c_2(F_R)\big)\\
&+ (\Lambda,\rho) \big(c_2(F_I)-c_2(F_F)\big)\ ,
\end{split}
\end{equation}
where $F_{L,R,I,F}$ are the background field strength for the $SU(2)_L \times SU(2)_F \times SU(2)_R \times SU(2)_I$ R-symmetry of the 2d $\mathcal{N}=(4,4)$ defect, and $c_2(F)$ is the second Chern class of the background $SU(2)$-symmetry bundle~\footnote{See~\cite{Drukker:2020atp} and the supplementary materials for the superconformal algebra with the surface defect insertion. }.

Since the surface defect is inserted in the 2d plane spanned by $\{x^0, x^1\}$ corresponding to $\omega_1$, the equivariant integral of the 2d defect anomaly polynomial \eqref{eq-2d-anomaly-polynomial} on this $\mathbb{R}^2_{\omega_1}$ gives \footnote{For details of the equivariant integration, see the supplementary materials.}
\begin{equation}
\begin{split}
 E_{\mathfrak{g}} 
 &= - \frac{1}{2\pi}\int I_4\\ 
 &= - \frac{1}{4\omega_1} \left[
   \frac{1}{2} \left(\Lambda,\Lambda\right) \left(a_L^2-a_R^2\right)+\left(\Lambda,\rho\right)\left(a_I^2-a_F^2\right)
   \right]\ ,
\end{split}
\end{equation}
where $a_L,a_R,a_I,a_F$ are the chemical potentials for the Cartans $J_L, J_R, J_I, J_F$ of the 2d $\mathcal{N}=(4,4)$ R-symmetry, respectively.
By identifying the 2d R-symmetry Cartans $J_L$, $J_R$, $J_I$, $J_F$ with the bulk Cartans $J_2$, $J_3$, $R_1$, $R_2$, the chemical potentials are given by~\footnote{see the supplementary material for the precise identification. },
\begin{equation}\label{eq-defect-chem-pot}
\begin{aligned}
    a_L &= \omega_2 + \omega_3\ , \qquad &a_R &= \omega_2 - \omega_3\ ,\\
    a_I &= - (\sigma_1 + \sigma_2)\ , \qquad &a_F &= - (\sigma_1 -\sigma_2)\ .
\end{aligned}
\end{equation}
Using \eqref{eq-defect-chem-pot} and \eqref{eq-b-d2-R}, the defect Casimir energy $ E_{\mathfrak{g}}$ can be further expressed in terms of the 6d bulk chemical potentials and defect central charges $b$, $d_2$ as
\begin{equation}
    \label{eq-Casimir-Final}
     E_{\mathfrak{g}} = -\frac{1}{\omega_1}\left[
    \frac{d_{2\mathfrak{g}} - b_{\mathfrak{g}}}{6} \omega_2 \omega_3 + \frac{2b_{\mathfrak{g}} - d_{2\mathfrak{g}}}{24} \sigma_1 \sigma_2 \right]\ .
\end{equation}
It can be checked that under the chiral algebra limit $ \omega_1 = \omega_2 = 1$, $\sigma_1 \sigma_2 = 2\omega_3$~\footnote{The surface defect should be orthogonal to the chiral plane \cite{Bullimore:2014upa, Chalabi:2020iie}. Since the surface defect is inserted on the $\mathbb{R}^2_{\omega_1}$ plane, here we choose $\mathbb{R}^2_{\omega_3}$ to be the chiral plane. The chiral limit under this choice is $\sigma_1 - \sigma_2 = (\omega_{1} + \omega_{2} - \omega_{3})/2$. We further take the limit $ \omega_1 = \omega_2 = 1$, and therefore $\sigma_1 \sigma_2 = 2\omega_3$. }, the defect Casimir energy \eqref{eq-Casimir-Final} is proportional to the defect central charge $d_2$~\footnote{It has been speculated that the entire defect supersymmetric Casimir energy is proportional to $d_2$~\cite{Chalabi:2020iie, Beccaria:2024gkq} but our result shows that this is only true in the chiral limit.}
\begin{equation}
E_{\mathfrak{g}}\big|_{\text{chiral limit}} = -\frac{1}{12} \omega_3 d_2\ .
\end{equation}
For the surface defect in the symmetric representation $(k)$ of $\mathfrak{g}=\mathfrak{su}(N)$, plugging the corresponding central charges \eqref{eq-b-d2-Sk} into \eqref{eq-Casimir-Final} gives
\begin{equation}\label{eq-anomaly-Sk}
E_{(k)} = -\frac{1}{2\omega_1}\left[ k^2 \left(1 - \frac{1}{N}\right) \omega_2 \omega_3 + k(N - 1) \sigma_1 \sigma_2 \right]\ .
\end{equation}
At the large $N$ limit, the anomaly polynomial-result \eqref{eq-anomaly-Sk} agrees with the localization-result \eqref{eq-logW-Sk}. One can also check this consistency for the anti-symmetric representation $[k]$. We stress that the defect SCE \eqref{eq-Casimir-Final} computed from anomaly polynomial is exact.

\textit{Defect SUSY entanglement entropy.} 
The value of defect SRE at $n=1$ is the surface defect contribution to SUSY EE. To define EE we have to specify the boundary condition at the entangling surface. For the defect contribution, the supersymmetric boundary condition gives a different result from the non-SUSY one (\ref{eq-JensenOBannon}). We assume that it is still a linear combination of $b$ and $d_2$, 
\begin{equation}\label{eq-susyEE-conj}
S = \log\langle D \rangle + \# d_2 \log\ell/\epsilon = \left(\frac{b}{3} + \# d_2\right)\log\ell/\epsilon\ ,
\end{equation}
where the undetermined constant $\#$ can be fixed to be $-1/6$ by fitting to the holography result \eqref{eq-holographyFinal} at $n = 1$. Therefore, the surface defect contribution to SUSY EE takes the following form~\footnote{Based on previous works~\cite{Huang:2014pda, Hama:2012bg, Alday:2009fs, Fucito:2016jng, Fiol:2015spa, Crossley:2014oea, Correa:2012at, Lewkowycz:2013laa, Fiol:2012sg}, we also verified a similar linear combination for the Wilson loop contribution to SUSY EE in 4d SYM theory, $S = \log\langle W \rangle - 6\pi^2 h_W$. See supplementary materials for details.}
\begin{equation}\label{eq-susyEE-surface}
S = \frac{2b -d_2}{6}\log\ell/\epsilon\ .
\end{equation}
There is an alternative way to verify the exact defect SRE \eqref{eq-main-result}. In~\cite{Yankielowicz:2017xkf} the authors observed an interesting simple relation between SCE and SRE for general even-dimensional SCFTs. Namely, extending the equivariant integration with an additional $U(1)$ and the chemical potential $\omega_4$ will give the exact SRE.
For the defect SCE \eqref{eq-Casimir-Final} in our case, the first term will not contribute SRE because it is linear to $n$ and the extension will not change the linearity. For the second term, plugging in $\sigma_1=\sigma_2=(3+1/n)/2$, $\omega_1=1/n$ and $\omega_{2,3,4}=1$, one can reproduce the defect SRE \eqref{eq-main-result} with $r_1=r_2=1/2$ precisely
 \begin{equation}
S_n = V_{\mathbb{H}^1}\frac{nE\big|_{n=1} - E\big|_n}{1-n} =  V_{\mathbb{H}^1}\frac{2b_{\mathfrak{g}} - d_{2\mathfrak{g}}}{12}\frac{1+7n}{8n}\ .
 \end{equation}

\textit{Defects in large $N$ limit.}
The $A_{N-1}$ $(2,0)$ SCFT is conjectured to be dual to M-theory on $\text{AdS}_7\times \mathbb{S}^4$ with $N$ units of 4-form flux on $\mathbb{S}^4$~\cite{Maldacena:1997re}. Furthermore, the holographic SRE of $A_{N-1}$ $(2,0)$ SCFT can be computed from 2-charge 7d topological black hole~\cite{Zhou:2015kaj},
\begin{equation}
\begin{split}
&\mathrm{d}s_7^2 = \frac{-f(r) \mathrm{d}t^2}{\left(H_1 H_2\right)^{4/5}} + \left(H_1 H_2\right)^{1/5} \left(\frac{\mathrm{d}r^2}{f(r)} + r^2 \mathrm{d}\Omega_{5,k}^2\right)\ ,\\
&f(r) = k - \frac{m}{r^4} + \frac{r^2}{L^2} H_1 H_2\ ,\quad H_i = 1 + \frac{q_i}{r^4}\ ,
\end{split}
\end{equation}
together with two scalars and two gauge fields,
\begin{equation}
X_i = \left(H_1 H_2\right)^{2/5}H_i^{-1}\ ,\quad A^i = \left(\sqrt{k} \left( H_i^{-1} - 1 \right) + \mu_i \right)\mathrm{d}t\ ,
\end{equation} 
where we consider $k=-1$ and $m=0$.
Let us define a rescaled charge $\kappa_i = q_i/r_H^4$, then the black hole horizon can be expressed in terms of $\kappa_i$,
\begin{equation}\label{eq-BHgorizon}
r_H = \frac{L}{\sqrt{(1 + \kappa_1)(1 + \kappa_2)}}\ .
\end{equation}
The vanishing condition of $A^i$ at the horizon fixes the chemical potential
\begin{equation}\label{eq-chemicalP}
\mu_i = \frac{i}{\kappa_i^{-1} + 1}\ .
\end{equation}
The thermodynamics of this black hole can be solved straightforwardly, and here we focus on solving a probe M2-brane, which is the holographic dual of the surface operator in fundamental representation.
The expectation value of a half BPS surface defect $-\log\langle D\rangle_n$ can be computed from the M2-brane on-shell action, $I_{\text{M2}} = T_2 \int \mathrm{d}^3\sigma \sqrt{-\mathrm{det}\left[g_{\text{ind}}\right]}$. A recent work~\cite{Yuan:2023oni} has shown that it only depends on the horizon,
\begin{equation}\label{eq-IM2}
I_{\text{M2}} = -\pi n T_2 V_{\mathbb{H}^1} r_H^2\ .
\end{equation}
To match the chemical potentials of the boundary CFT, one should set
\begin{equation}\label{eq-mui-ri}
\mu_1 = i(\gamma - 1)\frac{r_1}{2}\ , \quad \mu_2 = i(\gamma - 1)\frac{r_2}{2}\ , \quad \text{with}\quad r_1 + r_2 = 1\ .
\end{equation}
From \eqref{eq-BHgorizon}, \eqref{eq-chemicalP} and \eqref{eq-mui-ri}, $r_H$ can be expressed in terms of $r_i$ and $\gamma$,
\begin{equation}\label{eq-rH2}
r_H^2 = L^2 \left( 1 - (1 - \gamma)\frac{r_1}{2} \right) \left( 1 - (1 - \gamma)\frac{r_2}{2} \right)\ .
\end{equation} 
Using \eqref{eq-IM2} and \eqref{eq-rH2}, the defect contribution to holographic SRE is given by
\begin{equation}
\begin{split}
S_n &= \frac{\log\langle D\rangle_n - n \log\langle D\rangle_1}{1-n}\\
&= \pi T_2 V_{\mathbb{H}^1} L^2\big( r_1 r_2 (\gamma - 1) + 2 \big)/4\ .
\end{split}
\end{equation}
Given the holographic dictionary $N=L^3/(8\pi l_p^3)$ and $T_2 = 1/(4\pi^2l_p^3)$, it can be checked that when $n=1$, we have
$
I_{\text{M2}} = -4N\log\ell/\epsilon
$,
which agrees with the result in~\cite{Berenstein:1998ij, Drukker:2020dcz}. Moreover, we obtain the defect SRE in fundamental representation at the large $N$ limit, 
\begin{equation}\label{eq-holographyFinal}
S_\gamma = N \big( r_1 r_2 (\gamma - 1) + 2 \big)\log\ell/\epsilon\ .
\end{equation}

\textit{Discussions.} In this letter, we have proposed closed-form expressions for the surface defect contributions to the supersymmetric Rényi entropy and the supersymmetric Casimir energy in 6d (2,0) theories, formulated in terms of the defect central charges, $b$ and $d_2$. These formulas are supported by strong evidence and subjected to various consistency checks, further affirming their validity. Future directions include generalizing our approach to derive analogous formulas for co-dimension 2 defects, defects in 6d (1,0) SCFTs, and defects in 4d SCFTs. Additionally, it would be intriguing to explore bounds on central charges arising from these developments.

\textit{Acknowledgement.} We are grateful for useful discussions with our group members in Fudan University. This work is supported by NSFC grant 12375063. This work is also sponsored by Shanghai Talent Development Fund.

\appendix
\section*{Supplemental Material}
\subsection{Anti-symmetric representation}
Here we present the localization results for the anti-symmetric representation. The largest contribution of the trace in the anti-symmetric representation $[k]$,
\begin{equation}\label{eq-tr-Ak}
\mathrm{Tr}_{[k]} e^{2\pi\nu/\omega_1} = \sum_{1 \leq i_1 < \cdots < i_k \leq N} \exp\left[ \frac{2\pi}{\omega_1} \sum_{l = 1}^k \nu_{i_l}\right]\ ,
\end{equation}
comes with $\nu_{i_k} = \nu_N$, $\nu_{i_{k - 1}} = \nu_{N - 1}$, $\dots$, $\nu_{i_1} = \nu_{N - k + 1}$ \cite{Mori:2014tca}. Therefore the leading contribution to $\langle W_{[k]} \rangle$ is
\begin{widetext}
\begin{equation}\label{eq-matrix-Ak-WL}
\langle W_{[k]} \rangle = \frac{1}{\mathcal{Z}}\int \prod_{i = 1}^N \mathrm{d}\nu_i \exp\left[\frac{2\pi}{\omega_1 \omega_2 \omega_3}\left(-\frac{\pi}{\beta} \sum_{i = 1}^N \nu_i^2 + \frac{\sigma_1 \sigma_2}{2} \sum_{j < i}(\nu_i - \nu_j) + \omega_2 \omega_3 \sum_{i = N - k + 1}^N \nu_i\right)\right]\ .
\end{equation}
\end{widetext}
With this approximation, the Wilson loop insertion only changes the saddle points of $\nu_{i = N - k + 1, \dots, N}$, and the new saddle point equations are
\begin{equation}
- \frac{2\pi}{\beta} \nu_i + \frac{\sigma_1 \sigma_2}{2} \big((i - 1) - (N - i)\big) + \omega_2 \omega_3 = 0\ , 
\end{equation}
with the saddle solutions 
\begin{equation}\label{eq-vN-Ak}
\nu_{i = N - k + 1, \dots, N} = \frac{\beta}{4\pi} \big((2i - N - 1) \sigma_1 \sigma_2 + 2 \omega_2 \omega_3\big)\ .
\end{equation}
Evaluating the matrix integral \eqref{eq-matrix-Ak-WL} at the saddles \eqref{eq-vN-Ak}  of $\nu_{i = N - k + 1, \dots, N}$ and
\begin{equation}
\nu_{i < N - k + 1} = \frac{\beta \sigma_1 \sigma_2}{4\pi} (2i - N - 1)
\end{equation}
gives
\begin{equation}\label{eq-logW-Ak-SM}
-\log\langle W_{[k]} \rangle = -\frac{\beta}{2\omega_1} \big(k (N - k) \sigma_1 \sigma_2 + k \omega_2 \omega_3\big)\ , 
\end{equation}
At large $N$ limit,  \eqref{eq-logW-Ak-SM} agrees with the defect SCE calculated by the anomaly polynomial method,
\begin{equation}\label{eq-anomaly-Ak}
\begin{split}
E_{[k]} &= -\frac{1}{\omega_1}\left[\frac{d_{2[k]} - b_{[k]}}{6} \omega_2 \omega_3 + \frac{2b_{[k]} - d_{2[k]}}{24} \sigma_1 \sigma_2 \right]\\
&= -\frac{1}{2\omega_1}\left[ k \left(1-\frac{k}{N}\right) \omega_2 \omega_3 + k(N-k) \sigma_1 \sigma_2 \right]\ .
\end{split}
\end{equation}

\subsection{Superconformal algebra}\label{appx-SCalgebra}
\subsubsection{6d bulk  superconformal algebra}
The superconformal algebra of the 6d $(2,0)$ SCFT is $\mathfrak{osp}(8^*|4)$. Its maximal bosonic subalgebra $\mathfrak{so}(6,2) \oplus \mathfrak{so}(5)_R$ consists of the 6d conformal algebra $\mathfrak{so}(6,2)$~\footnote{Here the convention follows \cite{Drukker:2020atp}}
\begin{equation}\label{eqn:ospconformalalgebra} 
\begin{split}
&\left[ M_{\mu\nu}, M_{\rho \sigma} \right] = 2 \eta_{\sigma [\mu} M_{\nu] \rho} - 2 \eta_{\rho [\mu} M_{\nu] \sigma}\ ,\\
&\left[ M_{\mu \nu}, P_{\rho} \right] = 2 P_{[\mu} \eta_{\nu] \rho}\ ,\quad \left[ M_{\mu \nu}, K_\rho \right] = 2 K_{[\mu} \eta_{\nu] \rho}\ ,\\
&\left[ P_\mu, K_\nu \right] = 2 \left( M_{\mu\nu} + \eta_{\mu\nu} D \right)\ ,\\
&\left[ D, P_\mu \right] = P_\mu\ , \quad \left[ D, K_\mu \right] = -K_\mu\ ,
\end{split}
\end{equation}
and the $\mathfrak{so}(5)_R$ R-symmetry algebra
\begin{align}
  \begin{gathered}
  \left[ R_{IJ}, R_{KL} \right]
  = 2 \delta_{K [I} R_{J] L} - 2 \delta_{L [I}
  R_{J] K}\ ,
  \end{gathered}
\end{align}
where $\mu,\nu,\rho,\sigma=0,\dots,6$ are the indices of the 6d spacetime coordinates, and $I,J=1,\dots,5$ are the indices of the $\mathfrak{so}(5)_R$ R-symmetry vectors.
The 6d rotation group $\mathfrak{so}(6) \subset \mathfrak{so}(6,2)$ contains three Cartan generators $J_1,J_2,J_3$, while the 6d R-symmetry $\mathfrak{so}(5)_R$ has two Cartan generators $R_1,R_2$. These Cartans are related to the previous generators $M^{\mu\nu},R^{IJ}$ as
\begin{equation}
    \begin{aligned}
        &J_1 = i M^{01}\ ,\quad J_2 = i M^{23}\ , \quad J_3 = i M^{45}\ ,\\
        &R_1 = i R^{12}\ , \quad R_1 = i R^{34}\ .\\
    \end{aligned}
\end{equation}


\subsubsection{With the 2d surface defect insertion}
In the presence of the 2d $(4,4)$ surface defect, the original superconformal algebra $\mathfrak{osp}(8^*|4)$ is broken to $\mathfrak{osp}(4^*|2)\oplus \mathfrak{osp}(4^*|2)$, with the maximal bosonic subalgebra containing a 2d conformal algebra $\mathfrak{so}(2,2)$ along the defect plane, 
\begin{equation}
\begin{split}
&\left[ \mathsf{P}_\pm, \mathsf{K}_\pm \right] = 2 \mathsf{D}_\pm\ ,\quad \left[ \mathsf{D}_\pm, \mathsf{P}_\pm \right] = \mathsf{P}_\pm\ ,\\
&\left[ \mathsf{D}_\pm, \mathsf{K}_\pm \right] = -\mathsf{K}_\pm\ ,\\
\end{split}
\end{equation}
a rotation symmetry $\mathfrak{so}(4)_{\perp}\simeq \mathfrak{su}(2)_L \times \mathfrak{su}(2)_R$ on the remaining four transverse dimensions, 
\begin{equation}
\begin{split}
&\big[ \mathsf{T}_{(\perp)}^{\pm i}, \mathsf{T}_{(\perp)}^{\pm j} \big] = -i  \varepsilon^{ijk} \mathsf{T}_{(\perp)}^{\pm k}\ ,\\
\end{split}
\end{equation}
and a global symmetry $\mathfrak{so}(4)_R \simeq \mathfrak{su}(2)_I \times \mathfrak{su}(2)_F$ coming from the 6d bulk R-symmetry $\mathfrak{so}(5)_R$,
\begin{equation}
\begin{split}
&\big[ \mathsf{T}_{(R)}^{\pm i}, \mathsf{T}_{(R)}^{\pm j} \big]
  = -i  \varepsilon^{ijk} \mathsf{T}_{(R)}^{\pm j}\,, 
\end{split}
\end{equation} 
where  $i,j,k=1,2,3$, and ``$\pm$" denotes chiral and anti-chiral sectors, respectively.
Without loss of generality, we assume that the surface defect is inserted in the $\{x^0,x^1\}$-plane. The 2d conformal generators are then related to the 6d bulk ones as
\begin{equation}
\begin{split}
&\mathsf{P}_{\pm} = \frac{1}{2}(\mathsf{P}_0 \pm \mathsf{P}_1)\ , \quad \mathsf{D}_{\pm} = \frac{1}{2}(\mathsf{D}_0 \pm \mathsf{M}_{01})\ ,\\ 
&\mathsf{K}_{\pm} = \frac{1}{2}(-\mathsf{K}_0 \pm \mathsf{K}_1)\ ,
\end{split}
\end{equation}
and the generators of $\mathfrak{so}(4)_{\perp}$ and $\mathfrak{so}(4)_{R}$ are given by
\begin{equation}\label{eq-so(4)_gen}
\mathsf{T}_{(\perp)}^{\pm i} = \frac{i}{4} \eta^{\pm i}_{mn} M^{mn}\ , \quad \mathsf{T}_{(R)}^{\pm i} = -\frac{i}{4} \eta^{\pm i}_{ab} R^{ab}\ ,
\end{equation}
where 
\begin{equation}
    \eta^{\pm i}_{ab}= \pm (\delta^i_a \delta^4_b - \delta^i_b \delta^4_a) + \varepsilon_{iab}
\end{equation} 
are the 't Hooft symbols, $m,n=1,\dots 4$ are the indices of the four dimensions transverse to the defect plane~\footnote{Notice that the indices $m,n=1,\dots,4$ correspond to the 6d spacetime coordinates $\mu,\nu = 2,\dots,5$.}, and $a,b=1,\dots,4$ are the indices of the defect R-symmetry vectors.
\subsubsection{Identifying the defect R-symmetry Cartans}
The global symmetries $SO(4)_{\perp} \simeq SU(2)_L \times SU(2)_R$ and $SO(4)_R \simeq SU(2)_I \times SU(2)_F $ together form the R-symmetry of the 2d $(4,4)$ surface defect. More precisely, $SU(2)_L \times SU(2)_F$ is the left-moving R-symmetry, while $SU(2)_R \times SU(2)_I$ gives the right-moving R-symmetry. The Cartan generators of these four $SU(2)$ R-symmetry groups are
\begin{equation}
\begin{split}
    J_L &\equiv \mathsf{T}_{(\perp)}^{+3}\ , \qquad
    J_R \equiv \mathsf{T}_{(\perp)}^{-3}\ ,\\
    J_I &\equiv \mathsf{T}_{(R)}^{+3}\ , \qquad
    J_F \equiv \mathsf{T}_{(R)}^{-3}\ .\\
\end{split}
\end{equation}
With \eqref{eq-so(4)_gen}, they are related to the 6d bulk Cartans as
\begin{equation}
\label{eq-defect_Rsym_Cartan}
\begin{aligned}
    J_L &= \frac{1}{2}(J_2+ J_3)\ , \quad &J_R  &= \frac{1}{2}(J_2 - J_3)\ ,\\
    J_I &= -\frac{1}{2}(R_1+ R_2)\ , \quad &J_F  &= -\frac{1}{2}(R_1 - R_2)\ .\\
\end{aligned}
\end{equation}
Consider the pullback of the following quantity
\begin{equation}\label{eq-appxA-quantity}
    \omega_2 J_2 +\omega_3 J_3 + \sigma_1 R_1 + \sigma_2 R_2
\end{equation}
to the 2d defect. This quantity represents the 6d bulk global symmetries that are preserved by the 2d surface defect, which become the defect R-symmetry,
and therefore should be written in terms of the defect R-symmetry Cartans $J_{L,R,I,F}$ as 
\begin{equation}\label{eq-appxA-quantity-pullback}
\begin{split}
    &\omega_2 J_2 +\omega_3 J_3 + \sigma_1 R_1 + \sigma_2 R_2\\
    =\,& a_L J_L + a_R J_R +a_I J_I +a_F J_F\ ,
\end{split}
\end{equation}
where $a_{L,R,I,F}$ are the chemical potentials for $J_{L,R,I,F}$. Using the identification \eqref{eq-defect_Rsym_Cartan} of the 6d and 2d Cartan generators, \eqref{eq-appxA-quantity-pullback} then gives the relation between the 2d defect chemical potentials and the 6d bulk ones, 
\begin{equation}
\begin{aligned}
    a_L &= \omega_2 + \omega_3\ , \qquad &a_R &= \omega_2 - \omega_3\ ,\\
    a_I &= - (\sigma_1 + \sigma_2)\ , \qquad &a_F &= - (\sigma_1 -\sigma_2)\ .
\end{aligned}
\end{equation}
\subsection{Equivariant integral of the anomaly polynomial}\label{appx-eqv-int}
Here we use the Duistermaat-Heckman (DH) formula to evaluate the equivariant integral of the defect anomaly polynomial. The DH formula relates the equivariant integral of a $(d+2)$-form $I_{d+2}$ on $\mathbb{R}^{d}$ to the value of the $0$-form component of $I_{d+2}$ at the origin, 
\begin{equation}
    \frac{1}{(2\pi)^{d/2}}\int_{\mathbb{R}^{d}} I_{d+2} =  \frac{I_{d+2}|_0}{e(TM)|_0} \ ,
\end{equation}
where $I_{d+2}|_0$ and $e(TM)|_0$ are the 0-form component of $I_{d+2}$ and the equivariant Euler class at the origin, respectively. 
Using the DH formula, the equivariant integral of the 2d defect anomaly polynomial
\begin{equation}
\begin{split}
I_4 =&\,\frac{1}{2} (\Lambda,\Lambda)\big(c_2(F_L)-c_2(F_R)\big) \\
    &+ (\Lambda,\rho)\big(c_2(F_I)-c_2(F_F)\big)
\end{split}
\end{equation}
on the defect plane $\mathbb{R}^2_{\omega_1}$ leads to
\begin{equation}
 \frac{1}{2\pi}\int_{\mathbb{R}^{2}_{\omega_1}}\!\!\! I_{4} = \frac{I_4|_0}{e(TM)|_0}\ .
\end{equation}
At the origin, the $0$-form component of the equivariant characteristic classes $c_2(F)$ and $e(TM)$ are given by the equivariant parameters,
\begin{equation}
    c_2(F)|_0 =\frac{a^2}{4}\ , \qquad e(TM)|_0 = \omega_1\ ,
\end{equation}
and therefore we derive
\begin{equation}
    \frac{1}{2\pi}\int_{\mathbb{R}^{2}_{\omega_1}} I_{4} =
     \frac{1}{\omega_1} \left[
   \frac{1}{2} \left(\Lambda,\Lambda\right) \frac{a_L^2-a_R^2}{4}+
   \left(\Lambda,\rho\right)\frac{a_I^2-a_F^2}{4}
   \right] \ .
\end{equation}

\subsection{Supersymmetric Entanglement entropy (SUSY EE) for Wilson loops in 4d}\label{appx-WL}
Here we show that for 4d $\mathcal{N} = 4$ SYM theory, the 1/2 BPS Wilson loop contribution to SUSY EE is
\begin{equation}\label{eq-WL-SEE}
S = \log\langle W \rangle - 6\pi^2 h_W\ ,
\end{equation}
where $h_W$ is the coefficient in the bulk stress tensor one-point function in the presence of a straight Wilson line in flat space, i.e., 
\begin{equation}\label{eq-WL-h-def}
\llangle T_{00} \rrangle \equiv \frac{\langle T_{00}W\rangle}{\langle W\rangle} = \frac{h_W}{r^4}\ ,
\end{equation}
with $r$ the distance between $T$ and the Wilson line. $h_W$ is the analogy to the $d_2$ central charge of surface defects. One can see that \eqref{eq-WL-SEE} shows a similar structure as the surface defect SUSY EE. 

To derive \eqref{eq-WL-SEE}, recall that the R\'enyi entropy of a spherical region in CFT$_d$ can be calculated from the Euclidean path integral on a branched $d$-sphere $\mathbb{S}^d_n$~\cite{Casini:2011kv, Klebanov:2011uf}
\begin{equation}
\mathrm{d}s^2 = n^2\sin^2 \theta \mathrm{d}\tau^2 + \mathrm{d}\theta^2 + \cos^2 \theta \mathrm{d}\Omega_{d - 2}^2\ .
\end{equation}
Here $\tau \in [0,2\pi)$, $\theta \in [0, \pi/2]$, and $\Omega_{d - 2}$ is the standard $(d - 2)$-dimensional round sphere. In the following we will focus on $d = 4$. In the consideration of supersymmetry, the branched four-sphere $\mathbb{S}^4_n$ can be replaced by the squashed four-sphere $\mathbb{S}^4_b$ given by the equation~\cite{Huang:2014pda}
\begin{equation}\label{eq-ellipsoid}
\frac{x_0^2}{r^2} + \frac{x_1^2 + x_2^2}{\ell^2} +\frac{x_3^2 + x_4^2}{\tilde{\ell}^2} = 1\ ,
\end{equation}
with $\ell = nr$, $\tilde{\ell} = r$, and the squashing parameter is defined as $b\equiv (\ell/\tilde{\ell})^{1/2} = \sqrt{n}$. The partition function on $\mathbb{S}^4_b$ for arbitrary $\mathcal{N} = 2$ gauge theories has been computed using localization technique~\cite{Hama:2012bg}, and with an insertion of a BPS Wilson loop, the localization procedure also goes through. For example, in $\mathcal{N} = 2$ SCFTs, the expectation value of the Wilson loop operator on $\mathbb{S}^4_b$ is given by~\cite{Hama:2012bg} (see also~\cite{Alday:2009fs, Fucito:2016jng})
\begin{equation}\label{eq-matrix-WL}
\langle W_b \rangle = \frac{\int \mathrm{d}a\,e^{-\frac{8\pi^2}{g^2} \mathrm{Tr}a^2} Z_{\text{1-loop}}(a,b)|Z_{\text{inst}}(a,b)|^2 \mathrm{Tr}e^{-2\pi b a}}{\int \mathrm{d}a\,e^{-\frac{8\pi^2}{g^2} \mathrm{Tr}a^2} Z_{\text{1-loop}}(a,b)|Z_{\text{inst}}(a,b)|^2}\ ,
\end{equation}
with the Wilson loop located in the $\{x_1,x_2\}$-plane~\footnote{There are two BPS Wilson loops on the $\mathbb{S}^4_b$ \eqref{eq-ellipsoid} and they transform into each other under $\ell \leftrightarrow \tilde{\ell}$. The expectation value of the second Wilson loop can be obtain by replacing $\mathrm{Tr}e^{-2\pi b a}$ by $\mathrm{Tr}e^{-2\pi b^{-1} a}$ in the matrix integral~\eqref{eq-matrix-WL}. }. 

For $\mathcal{N} = 4$ SYM theory, near $b = 1$, the localization result simplifies to
\begin{equation}
\langle W_b \rangle = \frac{\int \mathrm{d}a\,e^{-\frac{8\pi^2 N}{\lambda} \mathrm{Tr}a^2} \mathrm{Tr}e^{-2\pi b a}}{\int \mathrm{d}a\,e^{-\frac{8\pi^2 N}{\lambda} \mathrm{Tr}a^2}} + \mathcal{O}\left((b - 1)^2\right)\ ,
\end{equation}
which is manifestly a function of $b\sqrt{\lambda}$ after the rescaling $a = \sqrt{\lambda}\tilde{a}$~\cite{Fiol:2015spa}, 
\begin{equation}
\langle W_b \rangle = \frac{\int \mathrm{d}\tilde{a}\,e^{-8\pi^2 N \mathrm{Tr} \tilde{a}^2} \mathrm{Tr}e^{-2\pi b\sqrt{\lambda} \tilde{a}}}{\int \mathrm{d}\tilde{a}\,e^{-8\pi^2 N \mathrm{Tr} \tilde{a}^2}} + \cdots\ .
\end{equation}

The Wilson loop contribution to SUSY EE is~\cite{Crossley:2014oea}
\begin{equation}
\begin{split}
S &= \left(1 - n\partial_n\right) \log\langle W_b \rangle |_{n \to 1}\\ 
&= \left(1 - \frac{1}{2}b\partial_b\right) \log\langle W_b \rangle |_{b \to 1}\ .
\end{split}
\end{equation}
Since $\langle W_b \rangle$ is a function of $b\sqrt{\lambda}$, there is a relation between its $b$-derivative and $\lambda$-derivative
\begin{equation}
\frac{1}{2} b\partial_b \log\langle W_b \rangle|_{b\to 1} = \lambda \partial_\lambda \log\langle W_b \rangle|_{b\to 1}\ .
\end{equation}
Using this relation, we obtain
\begin{equation}\label{eq-WL-susy-EE}
S = \left(1 - 
\lambda\partial_\lambda\right) \log\langle W \rangle\ ,
\end{equation}
with $\langle W\rangle \equiv \langle W_{b = 1}\rangle$ the expectation value of 1/2 BPS Wilson loop on a round four-sphere. 

For $\mathcal{N} = 4$ SYM theory, the $\lambda$-derivative of $\log\langle W\rangle$ determines the bremsstrahlung function~\cite{Correa:2012at}
\begin{equation}\label{eq-B-lambda}
B = \frac{1}{2\pi^2} \lambda \partial_\lambda \log \langle W\rangle\ , 
\end{equation}
and in~\cite{Lewkowycz:2013laa} the authors find that for $\mathcal{N} = 4$ theories, $B$ is related to $h_W$ by~\footnote{The relation \eqref{eq-B-lambda} was first noticed in \cite{Fiol:2012sg}. }
\begin{equation}\label{eq-B-h}
B = 3 h_W\ .
\end{equation}
With \eqref{eq-WL-susy-EE}, \eqref{eq-B-lambda} and \eqref{eq-B-h}, the Wilson loop contribution to SUSY EE can be finally expressed as
\begin{equation}\label{eq-WL-SUSY-EE-final}
S = \left(1 - \lambda\partial_\lambda\right) \log\langle W \rangle = \log\langle W \rangle - 6\pi^2 h_W\ .
\end{equation}
In the strong coupling regime $ \log\langle W \rangle\approx\sqrt\lambda$, this is consistent with the holographic result presented in~\cite{Crossley:2014oea}. Note that \eqref{eq-WL-SUSY-EE-final} is different from the Wilson loop non-SUSY EE~\cite{Lewkowycz:2013laa}
\begin{equation}\label{eq-WL-nonSUSY-EE}
\tilde{S} = \log\langle W \rangle + \int \llangle T^\tau_\tau \rrangle = \log\langle W \rangle - 8\pi^2 h_W\ .
\end{equation}

Here we give an explanation for the difference between the defect contribution to non-SUSY EE \eqref{eq-WL-nonSUSY-EE} and that for SUSY EE \eqref{eq-WL-SUSY-EE-final}. To illustrate, let us consider the calculation of \eqref{eq-WL-nonSUSY-EE} for free scalar theory. The second term proportional to $h_W$ comes from resolving the conical singularity at the entangling surface $\hat{\mit\Sigma}$ (the Wald term)~\cite{Lewkowycz:2013laa}. More precisely we smooth the tip of the cone and receive a $\delta$-function contribution to the Ricci scalar, which leads to the Wald term,
\begin{equation}\label{eq-WaldTerm}
S_{\text{Wald}} = -\frac{4\pi}{12}\int_{\hat{\mit\Sigma}}\mathrm{d}A \llangle\phi^2\rrangle\ .
\end{equation} 
A similar story occurs in 6d free scalar theory with Wilson surface insertion~\cite{Yuan:2023oni}.
However, in the case of SUSY localization, we did not follow the same boundary condition. Instead we choose alternative boundary conditions compatible with SUSY. This is the reason why we get a different coefficient in front of $h_W$  in \eqref{eq-WL-nonSUSY-EE} and \eqref{eq-WL-SUSY-EE-final}. 

\bibliography{biblio}
\end{document}